\documentstyle[12pt,epsfig]{article}
                      
\begin{document}
\pagestyle{myheadings} \markright{\bf On electromagnetic models of ball lightning }

\title{On electromagnetic models of ball lightning with topological structure}

\author{ 
 Jos\'e M. Donoso\\Departamento de F\'{\i}sica Aplicada,\\
ETSI Aeron\'auticos, Universidad Polit\'ecnica, 28040 Madrid,
Spain\thanks{E-mail: jmdonoso@aero.upm.es} \and
Antonio F. Ra\~nada\\ Departamento de F\'{\i}sica Aplicada III,\\
Universidad Complutense, 28040 Madrid, Spain \thanks{E-mail:
afr@fis.ucm.es} \and Jos\'e Luis Trueba\\ ESCET, Universidad Rey
Juan Carlos, 28933 M\'ostoles, Spain\thanks{e-mail:
jltrueba@escet.urjc.es}} 
\maketitle
\maketitle

\begin{abstract}
It has been long admitted that a consequence of the virial theorem
is that there can be no equilibrium configurations of a system of
charges in electromagnetic interaction in the absence of external
forces. However, recent results have shown that the virial theorem
can not preclude the existence of certain nontrivial equilibrium
configurations. Although some of these new results are based on an
effective microscopic field theory, they are important for a
theory of ball lightning that has been developed by the authors of
the present work.
 Other theoretical results relative to magnetic force-free fields
with field aligned currents and self-organized filamentary structures  
are also found to be relevant for this model. 

\end{abstract}

\section{Introduction}

Ball lightning is one of the oldest open problems of physics. It
consists in flaming balls or fireballs, usually bright white, red,
orange or yellow, which appear unexpectedly sometimes either near
ground, following the discharge of a lightning flash, or at midair
coming from a cloud.

Most of them are associated with thunderstorms, their shape is
usually spherical of spheroidal and they tend to move
horizontally. Typically, their diameter is between 10 and 50 cm.
The observed distribution of their lifetime has a maximum between
2 and 5 s and an average value of about 10 s or higher, some cases
of more than 1 minute having been reported. They are bright enough
to be clearly seen in daylight, the visible output being in the
range 10 to 150 W, similar to that of a home electric bulb (for a
detailed description of their properties see
\cite{Sin71,Bar80,Ste99}). Some balls have appeared within
aircraft, traveling inside the fuselage along the aisle from
front to rear. There are reports of odors, similar to those of
ozone, burning sulfur or nitric oxide, or of sounds, mainly
hisses, buzzes or flutters. The majority decay silently, others
with an explosion. They have killed or injured people and animals
and damaged trees, buildings, cars or electric equipment, fires
having been produced also. This shows that there is something hot
inside. In such events the released energy has been estimated
between 10 kJ and more than 1 MJ. They have never been produced in
laboratories, in spite of many attempts and some interesting
results, which have produced thus far some luminous objects that
decay too quickly. Consequently, their properties are investigated
on the basis of reports by witnesses, who are usually very
impressed and excited by the phenomenon and have no scientific
training.

A probably related phenomenon has been observed sometimes in
submarines, after a short circuit of the batteries: balls of
plasma have appeared at the electrodes and floated in air for
several seconds. In these occasions, the current was about 150 kA
and the energy was estimated between 200 and 400 kJ.

The most surprising property of a ball lightning is its great
stability and long lifetime, two related features which are very
difficult to explain. Indeed, nobody less than Faraday himself
stated in his {\it Experimental Researches in Electromagnetism}
that ball lightning can not be an electromagnetic phenomenon since
it would decay then instantaneously.

The previous features of the fireballs strongly suggest an
electromagnetic nature of the phenomenon. Note however that this
is compatible with the necessity of considering the chemical
aspects of plasmas, for instance in the determination of transport
coefficients such as the electric and thermal conductivities or
the viscosity.

There are two kinds of models of ball lightning, depending on
whether the energy source is external or internal. There has been
a tendency towards the internal energy models  since the
experimental verification, at the beginning of the nineties of the
existence of stable filamentary structures in a great variety of
plasmas, form laboratory to astrophysics. These observations
suggests that plasmas may have a self-organization ability that
could lead to long lifetime configurations. Electromagnetic models
with plasmas and magnetic fields since therefore  to be most
promising candidates to build a theory of BL.

However the fact remains that, in spite of many  attempts, there
is not an accepted theory of the energy confinement in fireballs,
specially of their long lifetime and great stability. It must be
added that some researchers interpret the magnetic virial theorem
as a serious obstacle to electromagnetic model since, according to
them, this theorem shows that the magnetic pressure would be so
strong that nothing could stand to avert an almost instantaneous
explosion \cite{Fin64}. This could be considered as a modern
formulation of Faraday's argument.

As a consequence, there has been a certain shift of the interest
of the researchers from purely or mainly electromagnetic to
chemical models. Aerosol model, for instance, have received
recently much attention.  In one of them \cite{Abr00}, the authors
argue that  a lightning discharge can vaporize silicon dioxide in
the soil which, after interacting with carbon compounds, is
transformed in pure silicon droplets of nanometer scale. These
droplets become coated with an insulating coat of oxides and are
polarized, after which they become aligned with electric fields
and form networks of filaments, in a loose structure that has been
called ``fluff balls". In another model \cite{Byc02}, the
discharges pick up organic material from the soil and transform it
in a kind of ``spongy ball" that can confine electric charges. It
seems difficult, however, to understand how these models could
explain the observed fact that some fireballs do appear near
clouds or in midair, where there is neither silicon nor organic
nor any other similar material.

We present in this paper some arguments, inspired both in
experimental and theoretical recent findings, which suggests that
this shift away from electromagnetic models may has been excessive
and that more emphasis must be put in them, even acknowledging the
necessity of paying due attention to the chemical properties of
the plasmas.

\section{Plasma model with a topological filamentary structure}
\subsection{The topological model}

In 1996, Ra\~nada and Trueba proposed a model with a topological
structure of magnetic lines and currents (``the topological model"
from now on), in which the fireballs are spheres of plasma coupled
to a magnetic field with non-vanishing magnetic helicity or,
equivalently, with linked magnetic lines and plasma streamlines
\cite{Ran96}. Although that model is very simple, it was useful to
show that the lifetime of such spheres increases with the linking
number of the lines. In other words tangled structures are more
stable. However, the plasma filled all the interior of the ball,
so that the radiation output was too high. A more elaborated model
was proposed by Ra\~nada, Soler and Trueba in 1998 \cite{Ran98}
and developed in 2000 \cite{Ran00}, in which the plasma is
contained in streamers, {\it i.e.} filamentary structures along
which the current flows, so that it fills only a small fraction of
the ball, of the order of one ppm of the volume. Moreover, the
streamers and the magnetic field are parallel. Such a
configuration is the result of the so called Taylor relaxation of
the plasma \cite{Tay86}, a very rapid process in which the
magnetic helicity is conserved, its value characterizing the final
relaxed state. Again a non-vanishing value of the linking number
gives stability to the system. What is important here is that the
ball decays rapidly to a quasi-force-free (FF) state, with finite
radius, and begins thereafter a slow expansion (termed {\it almost
quiescent} in \cite{Ran00}) which approaches asymptotically to a
force-free state.
 Such tangled structures are particular cases of
``electromagnetic knots", as were called in \cite{Ran90} the
fields with linked magnetic and electric lines.

In 2000, Faddeev and Niemi (FN) \cite{Fad00} presented an effective
microscopic field theory to give results on electrically neutral
plasmas with an equal number of negative and positive charge
carriers, which in their own words ``challenge certain widely held
views on plasma behavior". As already said before, it is
frequently admitted that the virial theorem implies that there can
be no equilibrium configurations of a system of charges in
electromagnetic interaction in the absence of external forces.
However, Faddeev and Niemi showed that this is not necessarily the
case, that simple arguments based on the virial theorem can not
preclude the existence of nontrivial, non-dissipative equilibrium
configurations, which are ``topologically stable solitons that
describe knotted and linked flux tubes of helical magnetic fields"
with ``self-confining plasma filaments". This results  is clearly
very important. It gives support to the topological model by
Ra\~nada, Soler and Trueba, which has precisely a structure of
linked currents and magnetic lines.

\subsection{The magnetic virial theorem}

Let us consider a plasma or gas cloud inside a volume $V$ bordered
by a surface $S$, with pressure $p$, density $\rho$, fluid
velocity {\bf v} and magnetic field {\bf B}. The momentum equation
can be written as
\begin{equation}
{\partial \over \partial t}(\rho v_i) +{\partial \over \partial x_k}(\rho v_iv_k) = -{\partial p\over \partial x_i}-\rho {\partial \Phi \over \partial x_i} +{1\over \mu _0}{\partial \over \partial x_k}\left[B_iB_k-{B^2\over 2}\delta _{ik}\right],
\label{0}
\end{equation}
where $\Phi$ is the self-gravitating potential, the last term being the Lorentz force ${\bf j}\times {\bf B}$, with ${\bf j}=\nabla \times {\bf B}/\mu _0$. Note that $B^2/2\mu_0$ acts as a pressure, usually called ``the magnetic pressure". It plays a similar role as $p$: if it has a great gradient it gives a large contribution to the force density. In many cases, it causes explosive expansions.

The magnetic virial theorem is the result of taking the first
momentum of this equation, i.e. of multiplying by $x_i$ and
integrating over the volume $V$ \cite{Chan87} (more precisely this
is the ``second order magnetic virial theorem". The ``first order
theorem" is just the integral of the Euler equation over the
volume $V$. It simply expresses the uniform motion of the center
of mass.) There is also a tensorial form of the theorem but it is
not relevant here. If we do that and assume that no fluid goes out
of the volume, i.e. that ${\bf v}\cdot {\bf n}=0$ if {\bf n} is  a
unit vector normal to the border $S$, we obtain the equation
\begin{equation}
{1\over 2} {d^2I\over dt^2} -2T =-\int _Sx_ipn_i dS +2U+W+U_B+{1\over \mu _0}\int _Sx_iM_{ik}n_kdS.
\label{1}
\end{equation}
where
\begin{eqnarray}
I &=&\int _V \rho r^2d^3x/2, \nonumber \\
T &=&\int _V\rho v^2d^3x/2, \nonumber \\
U &=&3\int _Vp\, d^3x/2, \label{1b} \\
U_B &=&\int _VB^2d^3x/2\mu _0, \nonumber \\
W &=&-\int _V{G\rho ({\bf r})\rho ({\bf r}^\prime )\over 2|{\bf r}-{\bf r}^\prime |}d^3x d^3x^\prime, \nonumber\\
M_{ik} &=& B_iB_k-B^2\delta _{ik}/2. \nonumber
\end{eqnarray}
$I$ is called the moment of inertia, $T$ is the kinetic energy, $U$ the internal energy, $U_B$ the magnetic energy, $W$ the gravitational self-energy. If the pressure is constant in $S$, $p=p_{\rm ext}$, the first integral in the RHS of (\ref{1}) is equal to $-3p_{\rm ext}V$. The gravitational self-energy is important for astrophysical plasmas but can be neglected in the case of ball lightning.

A frequent argument  states that the term $U_B$ in (\ref{1}) would produce a very rapid increase of $I$, {\em i.e.} an explosion, since nothing would balance the large magnetic pressure $B^2/2\mu_0$ (according to the witnesses reports, the magnetic energy of a BL must be of tens of kJ, even reaching several MJ). The calculations indicate that this would happen in a small fraction of a second, so that the virial theorem would exclude indeed any electromagnetic model of BL.

\subsection{Force-free magnetic knots and a model of ball
lightning}

However, Chandrasekhar and Woltjer showed in 1958 \cite{Cha58} that plasmas relax to minimum energy stable states, so called force-free fields (FFF), in which  the electric current and the magnetic field are parallel,
\begin{equation}
{\bf B}\times \left( {\mbox{\boldmath$\nabla$}} \times {\bf B} \right) =0,
\label{1c}
\end{equation}
so that the Lorentz force vanishes. They concluded that
such states can confine large amounts of magnetic energy.
If the force-free condition is verified, the sum of the two last terms in
 (\ref{1}) is zero. In these cases, the magnetic virial theorem reduces to
\begin{equation}
{1\over 2} {d^2I\over dt^2}= 2T - 3\int _V \delta p\, dV,
\label{40a}
\end{equation}
where $\delta p =p_{\rm ext}-p$ is the depression inside the ball,
i.e. the difference between the exterior and interior pressures.
As a consequence, the argument for an instantaneous explosion
fails (as happens in the topological model  \cite{Ran00} its
solution being dubbed ``force-free magnetic knots").

\subsection{ On filamentary BL plasma formation and dissipative processes}

In this section, we try to justify the existence of a filamentary
plasma structure by means of a more realistic  inspection into
plasma formation processes beyond the minimum energy relaxation
(under helicity conservation)
 arguments provided by Taylor.
 Otherwise, we are interested in investigate
how dissipative processes in a real plasma can be modified in a
magnetically dominated plasma regime to increase the lifetime of a
set of linked filaments.
 In the following, we shall refer to the
streamers, or a longitudinal section of one them,  as a subsystem
inside the global plasma ball structure (the system itself).

First of all, one has to be aware of our BL model
assumes an anisotropic plasma with linked streamers coupled
to the magnetic field in air.
For this reason, we provide here some arguments
related to the achievement of such filamentary structures that have been
actually observed
in natural plasmas. We are also concerned in looking for
 possible mechanisms to reduce
transport processes in a magnetically dominated plasma regime
since such processes would reduce plasma lifetime  due to energy
dissipation and charged particles diffusion. We think that both
problems, filamentary linked  plasma structure and
 dissipative processes reduction are directly related to
plasma dynamics driven by the magnetic field itself.

It is well known that filamentary and localized current-plasma
structures
 have been observed
experimentally in a wide range of plasma scenarios. From simple
 electrical discharges to Tokamaks and astrophysical plasma jets,
filamentation
 phenomena drastically lead the dynamics of plasma system, providing
 both positive
 and negative effects on the global stability. Otherwise, due to plasma
 ability to self-organize,
 one can expect the existence of a meta-stable system if some
 conditions are really fulfilled.

For instance, large stable filamentary currents in plasma may last
in a larger time scale compared to plasma energy confinement time.
Since this property has been experimentally tested in several
plasma regimes, as predicted theoretically by Faddeev and Niemi in
\cite{Fad00} , a plasma system mainly composed of such filamentary
long-living structures could last a long time. Note that in the
atmosphere, the fireball may be also surrounded by a partially
ionized plasma air, at last in the first stages before reaching a
stable structure. We mean here, that it would be possible to
combine a fully ionized plasma regime inside each streamer with a
surrounding partially ionized cold plasma in our model. This
property could be very important if one deals with eventual energy
sources in each current channel.

In treating on filamentary plasma formation,
let us assume that in an ordinary lightning a vortex type structure of
plasma can be reached in atmospheric conditions.
 Thus, a toroidal like plasma could exist encircling the straight
  discharge channel of an ordinary
lightning following the strong magnetic field associated with the
lightning current.
 This primary toroidal plasma could also be surrounded by secondary
current channels and all of then could be filled increasingly by
charge injection coming from the main stroke that
 branches, and returns,  through
partially ionized channels. There are some photographic evidences
of such lightning formations where it can be seen a tangled
structure, sometimes not a single one, along lightning path.
Moreover, this tangled lightnings have been also seen associated
to a ball lightning\footnote{\scriptsize

$
\mbox{http:}//\mbox{www.ernmphotography.com}/\mbox{Pages}
/\mbox{Ball\_Lightning}/\mbox{Ball\_Lightning\_ErnM.html}$

}.

 At this stage, it can be assumed that
 the strong lightning current source
suddenly disappears along the vortex axis so that
 large pressure gradient can induce
a plasma compression. The confining magnetic field disappears also
and secondary currents channels start to self-organize into a new
system. Note that due to the different mobilities of electron and
(massive) ions, charge separation occurs after the lightning
decay. This happens in a time scale shorter than the one required
to destroy completely the secondary configuration, so that the
current has time enough to collapse before the vortex destruction
into a set of thin channels having relatively large energy
densities. This collapse may be achieved by attraction of laminar
currents carrying opposite charges. A simpler explanation for the streamer
linked structure can also be given by just considering  
self-cutting (by pinch effect in both incoming and outgoing extremes) 
and separation of the 
ordinary lightning tangled formation followed by current collapse and 
compression.

Note that this mechanism would
 explain
 the existence of closed streamers created by simple charge collapse, not by
the usual single sign charge  avalanche in ordinary arc discharges. Otherwise,
 the plasma can also rotate as well as it can
 modify its structure due to air whirlwind enhancing the tangled or eddy for
of the new plasma regime while magnetic reconnection starts. We
mention here that recent laboratory experiments show that a plasma
ball with loop formations can be ejected from a burning metallic
wire.

As said before, toroidal like  fluid flow rotating tube
structures in the initial vortex will be compressed by simple
collapse,  providing twisting  and filamentation in the charge
distribution currents in an almost instantaneous magnetic
reconnection process.
 The role played by finite thermal conductivity and Lorentz forces are
relevant to energy conversion into  the new  self-organized  state.
The result may be a set of relatively high density
linked streamers or filaments strongly  coupled to the structure of
the new magnetic field.

Up to now we have argued in favor of a {\bf meta-stable
self-organized } system of charged particles  coupled to a
magnetic force-free field, but concentrated only on the standard
Taylor relaxation processes. However it has been recently shown
that a plasma can relax to a Taylor force-free state under more
general assumptions than helicity conservation constraints. In
this sense, Zhu et al
  \cite{zhu} have shown that a magnetohydrodynamic plasma can
self-organize to a force-free or quasi-force-free  state
 if
finite thermal conductivity is taken into account during magnetic
reconnection, while a non-Taylor state with electric current
perpendicular to magnetic field is reached without thermal
conduction. The enhanced heat conduction tends to smooth
nonuniformity of both peaked pressure and perpendicular electric
current component during driven magnetic reconnection. In this
way, the Taylor force-free state under no pressure gradients is
approximately realized in a short time (microseconds). The role
played by   magnetic reconnection combined with heat conduction
could be very important to explain field-aligned currents and
magnetic energy conversion into kinetic energy. As shown in
\cite{yoko}, the effect of magnetic reconnection, enhanced by heat
conduction along field lines, can explain the loop structures in
solar chromosphere. Note that our ball lightning model
(\cite{Ran98,Ran00}) could share some traits with astrophysical
phenomena, because of the MHD scale invariance.

From our point of view, the above  argument is relevant to the
formation of almost relaxed states, which are very close to
force-free configurations in air. Furthermore, it explains the
direct relation of the current plasma alignment and the subsequent
current-field coupling, as well as the direct heating of the
particles and the high deposition of energy into the system. Otherwise,
it should be noted that magnetic reconnection plays also an important role 
in redistributing total magnetic helicity, globally conserved, providing
locally twisting of currents filaments and contributing to global plasma
stability after a self-organized state is reached.

Moreover, it is well-known that the magnetic reconnection
generates {\bf charge acceleration  and system heating}, because
of this, the current ${\bf J}$  and magnetic field strength $B$
 inside each filament can be very large. This is an important feature because
of large magnetic field  generation and current alignment may be
very important to justify the long lifetime of a fireball.

It is remarkable that one can not reject the possibility
 of certain large number of
 suprathermal electrons or runaways, as  experimentally found in upper
atmosphere storms.
Note that
 a nonvanishing  suprathermal electrons population
would increase the plasma electrical conductivity $\sigma $
providing a  highly conductor fluid  slowing down the
 magnetic helicity decay rate,
$\eta \int {\bf J} \cdot {\bf B} d^3 {\bf r} ,$  keeping the field topology,
if one assumes a real plasma with finite
resistivity $\eta .$

These arguments lead us to treat with
some non-ordinary  plasma regimes  inside each channel
subsystem as well as inside the whole anisotropic plasma ball.

 Roughly speaking, the magnetic field inside a streamer may be treated as a
singularity in the global field, although its size (radius) may be
much greater than the standard Debye length and typical plasma
skin depth. Under such assumption, we have tested the  evolution
time  of an isolated plasma column treated as  a longitudinal
cylindrical filament. Mechanical effects leading to energy losses
can be neglected in a force-free state because of this energy
decay is given by the volume integral
\begin{equation}
 \epsilon _{mech.} = \int {\bf v} \cdot ({\bf J} \times {\bf B}) d {\bf r}
 \end{equation}
which vanishes exactly inside any streamer.
Otherwise, magnetic field diffusion, can be reduced if the new evolutionary 
plasma system departs from a force-free state with field-aligned fluid velocity
since 
\begin{equation} 
\frac{\partial{\bf B}}{\partial t} = \nabla \times ( {\bf v} \times {\bf B}) 
+ \frac{ 1}{ \mu \sigma} \nabla ^2 {\bf B}  \label{difu}
\end{equation}
the first term vanishes, and $\nabla ^2 {\bf B} $ $= - \lambda ^2{\bf B}$
for a linear
 force-free field having $\nabla \times {\bf B} = \lambda {\bf B}.$ 
Thus, the magnetic field strength $B$ would decay exponentially with a 
relaxation time $ \tau = \mu \sigma / \lambda ^2 $ which increases linearly
with conductivity $\sigma$ and could be very large if we associate 
$1/\lambda$ to a system  characteristic length  $L$ and we take for $L$ 
the length of a streamer that may be several meters in lenght.   

At this stage, 
the question relative to justify the stability and long lifetime  
 of our system arises naturally. 
Since exploring quantitatively the stability of the configuration is still a
subject for future works,  
we only treat here  with the problem of ordinary dissipative processes. 
 In an isolated 
simple plasma regime  without energy sources,  thermal transport will lead to
an energy decay time  
for a plasma  in a range  $50-100 \mu m $  for  
streamer's  width of a fraction of second. 
In spite of local (streamer) and global (plasma ball) 
stability  is a direct consequence of the strong 
linkage between the filaments, in this qualitative approach on for an isolated
plasma column 
we must keep in mind that the severe anisotropic structure provided by 
the magnetic field can also contribute to increase  the plasma lifetime,
even for a relatively small subsystem which is a current filament.

Using the time evolution equation for the
  fluid enthalpy volume density   ($h=\rho c_p T ,$) in an air plasma
\begin{equation}
\label{fluxh}
\rho \frac{d h }{d t} =   - \nabla {\bf q} + S \
\end{equation}
 with mass density $\rho$,
 mass specific heat at constant pressure $c_p$ and  heat flux ${\bf q}$ we
deal with time dependence of kinetic temperature radial profile $T(r)$.
 As
the filament is supposed to be closed,   above equation only takes
into account perpendicular heat flux transversal to ${\bf B}$
i.e., parallel heat conduction does not contribute effectively to
dissipation processes because of the closed nature of the
filament. Moreover, particles can flow almost freely along the
field  in a quasi collision-less regime. Note that the mechanical
energy loss rate $\int {\bf v} \cdot ({\bf J}\times {\bf B}) dV$
vanishes identically if
 a force-free configuration inside the column is assumed.
The heat flux reads ${\bf q}= -  \lambda _{\perp}  \nabla _{\perp}
T ,$ after  dropping other energy sinks or sources $S$. Here, $
\lambda _{\perp} $ is the perpendicular component of the thermal
conductivity tensor which has been modeled assuming
 classical  behavior   \cite{brag}:
$\kappa _{\perp} \simeq $ $ \kappa_{\parallel}
/ \omega ^2 \tau ^2 $ $ \sim 1/(B^2 T^{1/2}) $ for
thermal conductivity coefficients.
 In this sense, the thermal isotropic conductivity
$\lambda _0=\kappa m c_p$ for a reacting plasma in air having mean
molar mass  $m=m(T)$, taken from the work of Capitelli et al in
\cite{capitelli}, is modified by a factor $\epsilon =1/(1+ \omega
^2 \tau ^2) $ providing  $ \lambda _{\perp}=\lambda_0 \epsilon.$
Here, $\omega= eB/m$ is the electron gyro-frequency  and $\tau$ is
the collision time,
 obtained from electrical standard conductivity in an air plasma.
The simple force-free magnetic field
 ${\bf B(r)} = ( 0, B_{\phi}(r),B_z(r)) $ in cylindrical
polar coordinates
is assumed to be constant in time with $B_{\phi}$ increasing from zero to a
maximum
value $B$ at $r=a$. The functional dependence on $r$ is not so longer relevant
for our actual aims.

 As a departure state we consider a smooth radial
temperature profile decaying to ambient temperature (over $1,000K$
in a surrounding partially ionized air). Since there is no
physical nozzle  enclosing the current, free boundary conditions
are considered, choosing as streamer effective radius $a$ the
distance at which plasma temperature is  about $14,000K.$   Note
that under these assumptions, the magnetic field may provide the
subsystem by an effective confinement
 magnetic nozzle. In this sense, for instance,
a similar  kind of plasma regime  has been recently investigated
 in \cite{wang}. In this work,  the authors  present a plasma regime
that can be confined without a physical nozzle
(magnetic nozzle system)
if the plasma is collisionless only along the magnetic field.

\begin{figure}[htb]
\vskip -11pt
 \begin{center}
 \setlength{\unitlength}{1.0cm}
  \mbox{\epsfig{file=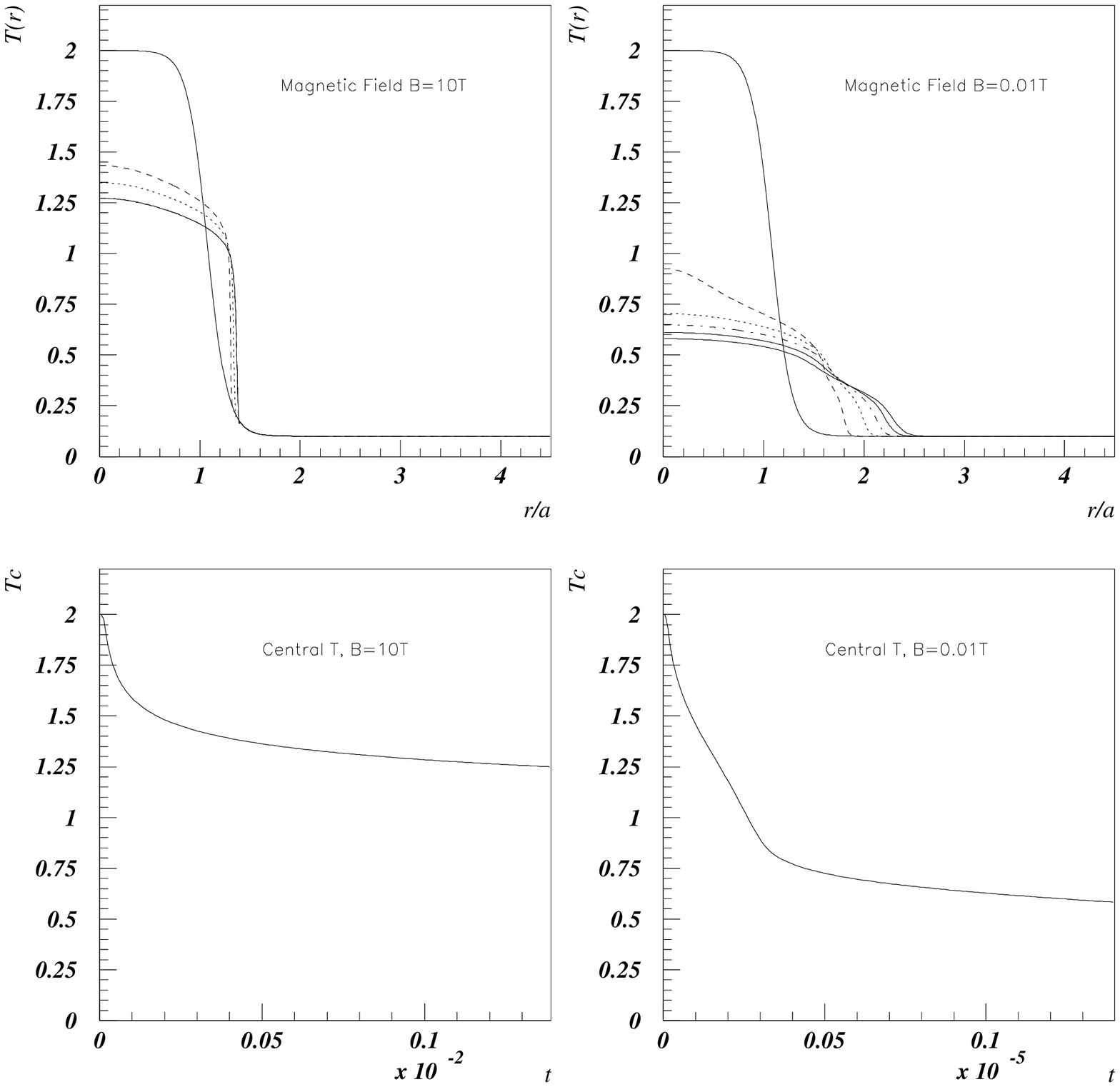,width=12.5cm}}
\end{center}

\scriptsize
\caption{ \ \ \em Kinetic Temperature profiles inside a filament  }
{ \smallskip
 \bf Radial temperature streamer profile in $10^4$K.
Under  a strong magnetic field the free  streamer radius
tend to be constant and  diffusion expansion is  drastically reduced changing
the characteristics
 temporal  scale. Time evolution of central temperature
$Tc(t),$ ($t$ in s)  at $r=0$ is also shown.
Clearly, the system
 exhibits a large  time scale evolution for magnetized filament.
The slope for  almost linear decreasing $Tc(t)$ is greater under
small or vanishing magnetic field.  } \label{fig1}
\end{figure}

As shown in figure \ref{fig1}, temperature diffusion implying
filament destruction is clearly inhibited by  a strong enough
local magnetic field.  This means that transversal heat transport
due to electrons is  reduced in a strongly
magnetized media. Dissipative processes would take place in a
larger temporal scale if it is compared to similar diffusive
situations for the isotropic case (low or null magnetic field).
Note that the temperature $T_c$ at $r=0$ also decreases slowly
when this value reaches approximately $1,600K$, around this value,
the air thermal conductivity also shows a minimum value leading
the system to be in an almost stable state with low thermal
dissipation.

\section{Conclusions}

It can be said that the main problem for the plausibility of an electromagnetic model of BL is to reconcile the interpretation of the classical virial theorem with the recent arguments on electrically neutral plasmas presented by Faddeev and Niemi \cite{Fad00}. In this work, we have related the FN results with a model of ball lightning \cite{Ran00,Ran98,Ran96} that uses the concept of force-free magnetic knot. The conclusion is that a frequent argument against electromagnetic models of ball lightning fails.

There are two other reasons why the FN result is interesting. One is that the power emitted by a plasma ball  of the size of a BL is too high (one litre at 15,000-20,000 K emits about 5 MW), which is an indication that most of the ball is cold, only a fraction being concentrated in filamentary structures as hot current streamers as in reference \cite{Ran98}. Another: it is sometimes assumed as a matter of fact that the streamers can not last because they will be necked or cut in a very short time by the pinch effect ({\em i.e.} by the Lorentz force), contrary to the FN result. But this does not happen if the FFF holds, since there is no pinch effect, so that the streamers would be stable. In the case of almost FFF, the pinch effect would be small and would contribute only weakly the decay, which will be produced only after a certain much longer time, another
stabilizing factor.

Force-free  states can be reached not only under Taylor
assumptions, such states may also appear in Nature as a
consequence of the implication of dissipative processes as thermal
conduction during (and after) magnetic reconnection that also 
plays a significant role in field-current aligned structure, 
charge acceleration and redistribution of global helicity to improve stability
and  self-organized quasi-force-free state. 
The ordinary  dissipative processes are drastically reduced due to anisotropic
structure in the system induced by the magnetic field.Thus,  a
self-organized  plasma structure through charge and
 currents concentration in linked filaments can  provide a stable long living
plasma system with reduced dissipation.

\newpage

\end{document}